\documentclass[prl,showpacs]{revtex4}
\usepackage{graphicx}
\usepackage{amsmath}
\usepackage{amssymb}
\usepackage{color}
\usepackage[normalem]{ulem}
\usepackage{epsfig}

\begin{document}
\title{Age-abundance relationships for neutral communities}
\author{Matan Danino and Nadav M. Shnerb}
\affiliation{Department of Physics, Bar-Ilan University, Ramat-Gan
IL52900, Israel} 
\begin{abstract}
Neutral models for the dynamics of a system of competing species are
used, nowadays, to describe a wide variety of empirical communities. These models are
used in many situations, ranging from
population genetics and ecological biodiversity to
macroevolution and cancer tumors. One of the main issues discussed
within this framework is the relationships between the abundance of
a species and its age. Here we provide a comprehensive analysis of
the age-abundance relationships for fixed-size and growing communities. Explicit formulas for the average
and the most likely age of a species with abundance $n$ are given,
together with the full probability distribution function. We further discuss the universality of these results and
their applicability to the tropical forest community.
\end{abstract}
\maketitle

\section{Introduction}

An understanding of the mechanisms underlying the composition of a
community of competing species is one of the important problems of
contemporary science. Ecologists would like to understand what
determines the stability of high-diversity assemblages like
freshwater plankton \cite{hutchinson1961paradox,stomp2011large} or trees in the tropical forest \cite{ter2013hyperdominance}, geneticists are
interested in the forces that govern genetic diversity within a
population, other researchers would like to explain the dynamics of
different clones within a malignant tumor \cite{Sottoriva2015cancer} or the patterns of
macroevolution \cite{maruvka2013model}. All these systems are characterized by competition
between many "species" (clones, haplotype etc.) with an underlying
birth-death process that allows for abundance variations and is combined
with the possibility of mutation/speciation by which new species are
originated.

Traditionally, the attempts to explain the dynamics of such
communities were based on "niche" approach, i.e., on the
identification of a special feature of any species (like better
efficiency in the consumption of a certain resource etc.), a feature
that provides an advantage for this species and limits the strength of  its
competition with other species \cite{chesson2000mechanisms}. This
approach poses many conceptual problems, like the
identification of so many different niches
\cite{hutchinson1961paradox,grubb1977maintenance} and the stability
of a complex community with many competing species
\cite{may1972will}. These problems, together with the practical difficulty in
the calibration of parameters for high-dimensional models,
\cite{connolly2014commonness}, have motivated the search for an
alternative paradigm.

During the last decades a radical alternative to the concept of
niche-based analysis was presented in various fields. \emph{Neutral} theories
\cite{kimura1985neutral,hubbell2001unified} suggest a new model,
under which (at least as a first order approximation) all individuals  in a community are
competing equally for a
single resource (e.g., space). To be more specific, there is no
niche separation (the competition between any two individuals is the
same, independent of their species identity) and the fitness of
all individuals is, again, the same (demographic equivalence). In a
neutral game, the outcome, like the abundance of a species or its
chance to perish, depends solely on the stochastic processes in the
system, and in particular on the demographic fluctuations that
characterize the birth-death-mutation dynamics.

An important distinction between various types of neutral theories
has to do with the overall size of the community. In ecology and
genetics the research is focused on communities that have, more or
less, a fixed size, assuming that the overall biomass of the
community have already reached a stable equilibrium. On the other
hand, when the neutral theory is applied to macroevolution
\cite{maruvka2013model}, to the dynamics of networks
\cite{lansky2013role,maruvka2011birth} or to surnames in human population
\cite{rossi2013surname,maruvka2010universal,manrubia2002boundary}, one has in
mind an exponentially \emph{growing} community. Recently, a few
authors applied the neutral theory to a growing malignant tumor,
trying to understand intra-tumor heterogeneity and mutant allele
frequencies \cite{Sottoriva2015cancer}. In particular, the
heterogenous response of individuals to drug therapy as was
interpreted as reflecting the abundance of resistent cells, assuming that sensitive and resistant clones
have the same demographic parameters before the treatment
\cite{antal2011exact,iwasa2006evolution,kessler2014resistance}.

In many cases the success of the neutral theory in explaining
community composition (especially species abundance distribution (SAD)) is quite impressive, given its minimalistic
assumptions and the small number of parameters involved
\cite{Volkov2003Neutral,maruvka2010universal}.  However, the predictions of the neutral
theory should be examined also in relation to the  \emph{dynamics} of empirical systems \cite{kalyuzhny2014generalized}.
In this context, one of the main differences between neutral and niche theories has to do with the relationships between the
age of a species
(the number of generations elapsed since the speciation event, i.e., the birth of the founder) and its abundance.
In niche theories there is a very weak correlation between these two quantities: the model supports an
 attractive fixed point that depends on the relative fitness of the competing species, and the equilibration time
 is almost independent of the number of individuals. Under neutral dynamics, on the other hand, the system is governed by stochasticity.
 Accordingly, in a fixed size community the time needed for a single founder to establish population of size $N$ is typically $N$
 generations, and in a growing community the age of a species is positively correlated with its age.
Indeed, one of the
 known difficulties of  Hubbell's neutral  theory of biodiversity, has to do with the age-abundance relationships,
 since the $N$ generations requirement appears to be unrealistically long
 \cite{nee2005neutral,ricklefs2006unified,chisholm2014species}. For growing communities, like in the case of macroevolution,
 one can find contradicting statements about the correlation between taxon's
 age and its size \cite{rabosky2012clade,mcpeek2007clade,maruvka2013model}.

This paper is dedicated to the age-abundance problem under neutral dynamics. We
 extend the results of a recent
work by Chisholm and O'Dwyer \cite{chisholm2014species}, who provide an analytic expressions for these relationships in a fixed size
community.  Here we consider also growing (or shrinking) communities, showing that under neutral dynamics, the species
abundance distribution determines unequivocally the relationships between abundance and mean age of a species. We further
 explore the conditions under which these results are universal, i.e., independent of the details of the process (like
 overlapping/non-overlapping generations, or the statistics of offspring per individual).

This paper is organized as follows: in the first few sections we present
a solvable  model for Wright-Fisher dynamics (non-overlapping generations) with geometric
distribution of offspring. $P_n^{(s)}$, the chance of a species to have abundance $n$, $s$ generations after its origination
is calculated, along with the maximum likelihood expression for the age of the species, $s$,  given $n$. We then proceed
to calculate $T^*(n)$, the average age of a species given its abundance, showing that this quantity is determined by
the species abundance
distribution. Finally, we discuss the universality of our results and its applicability to the
empirical findings in the Amazon forest.

\section{A solvable model}

The master equation for a \emph{continuous time} neutral dynamics of
a single population with pure demographic noise (a Moran process) is
given by,
\begin{equation} \label{eq1a}
\frac{dP_m}{dt} = -2mP_m+(m+1)P_{m+1} + (m-1)P_{m-1},
\end{equation}
where $P_m$ is the chance that the species is represented by $m$
individuals. This equation corresponds to a process in which every
individual tosses a coin at random, dying if the outcome is tails and
producing an offspring when the outcome is a heads. The probability
(per individual per unit time) for such an event sets the timescale
in Eq. (\ref{eq1a}). Accordingly, the chance that an individual
produces $n$ offspring during its lifetime (from birth to death) is
given by a geometric distribution, $P^{(1)}_n = 1/2^{n+1}$, where
the superscript $(1)$ indicates that this is the probability to
produce $n$ offspring during a single ''generation".

Here we  consider a variant of this model, with the same offspring
statistics but with non-overlapping generations (Wright-Fisher
dynamics). In this version at every generation each individual
produces $n$ offspring with a chance $P_n = 2^{-(n+1)}$ and dies.
The only difference between the continuous and the discrete time
versions is the definition of a generation, and as we shall see
below in the relevant regime of parameters this modification has no
importance.

For the geometric distribution it is quite easy to calculate the one
generation generating function,
\begin{equation} \label{eq2}
G^{(1)}(x) = \sum_{n=0}^{\infty}x^n P^{(1)}_n = \frac{1}{2-x},
\end{equation}
and the  structure of this generating function
\cite{steffensen1933} allows for an explicit formula for the
generating function  after $s$ generations, which is obtained by
iterating $G^{(1)}$ $s$ times,
\begin{eqnarray} \label{eq3}
G^{(s)}(x) &\equiv& \sum_{n=0}^{\infty}x^n P^{(s)}_n = G(G(G...\rm{s
\  times} \ (G(x)))) \\ \nonumber &=& \frac{s-(s-1)x}{(s+1)-sx}.
\end{eqnarray}

The probability of a single individual to produce $n$ offspring
after $s$ generations, $P^{(s)}_n$, is the coefficient of $x^n$ in
the series expansion of $G^{(s)}(x) $. This yields (for $n
> 0$),
\begin{eqnarray} \label{eq4}
P^{(s)}_n = \frac{1}{(s+1)^2} \left(1-\frac{1}{s+1}\right)^{n-1}
 \approx \frac{1}{(s+1)^2} e^{-\frac{n-1}{s+1}},
\end{eqnarray}
while for $n=0$,  $P^{(s)}_0 = s/(s+1)$. Accordingly, the probability
of survival $s$ generations decays in long times like $1/s$, in agreement
with the general results of Galton-Watson theory \cite{galton1874}.

To generalize this model we consider a case where the average number
of offspring produced by a single individual is $1+\gamma$ (so
$\gamma$ is the growth/decay rate of the population); the chance of success in the
corresponding  Bernoulli trial should be $z=(\gamma+1)/(\gamma+2)$. In addition we consider a process with
mutation/speciation events: the chance that an offspring
does not belong to the same taxon of its mother is $\nu$ (no recurrent mutations, and mutant is the founder of
a new species).  For this birth-death-mutation process the
average number of non-mutant offspring per individual is $R_o \equiv 1+r =
(1+\gamma)(1-\nu)$.

For the Bernoulli process considered here the probability for $n$
non-mutant offspring is given by:
\begin{eqnarray} \label{eq6}
P^{(1)}_n &=& \left(1-z \right) \sum_{k=n}^{\infty} z^k \binom{k}{n}
(1-\nu)^n \nu^{k-n} \\ \nonumber &=&\frac{R_o^n}{(1+R_o)^{n+1}},
\end{eqnarray}
 hence the generating function is,
\begin{equation} \label{eq7}
G^{(1)}(x) =  \frac{1}{1+R_o-R_o x}.
\end{equation}
Although the recursion relation is less trivial, one can still
derive the generating function after $s$ generations. To do that, it
is useful to notice that during the recursion process (plugging
$G^{(1)}$ instead of $x$ in (\ref{eq7}) to get $G^{(2)}$ and so on)
$G$ retains its general form,
\begin{equation} \label{eq8}
G =  \frac{\alpha_s + \beta_s x}{\gamma_s + \delta_s x},
\end{equation}
and the recursion from the $s$ to the $s+1$ generation satisfies,
\begin{equation} \label{eq9}
\left[ \begin{array}{c} \alpha_{s+1} \\ \beta_{s+1} \\ \gamma_{s+1} \\
\delta_{s+1}
\end{array} \right] =
\begin{bmatrix} 0 & 0 & 1 &0  \\ 0 & 0 & 0 & 1 \\ -R_o & 0 & 1+R_o & 0 \\ 0 & -R_o & 0 & 1+R_o \end{bmatrix}  \left[
\begin{array}{c} \alpha_{s} \\ \beta_{s} \\ \gamma_{s} \\ \delta_{s} \end{array}
\right],
\end{equation}
with the initial conditions   $\alpha_1 = 1, \ \beta_1 = 0, \
\gamma_1 = 1+R_o$ and $\delta_1 = -R_o$. Accordingly,
\begin{eqnarray} \label{eq10}
G^{(s)}(x)
=\frac{(R_o^s-1)-(R_o^s-R_o)x}{R_o^{s+1}-1+x[R_o(1-R_o^s)]},
\end{eqnarray}
and,
\begin{eqnarray}  \label{eq19}
P^{(s)}_n &=& (1-R_o)^2 R_o^{s+n-1}
\frac{(1-R_o^s)^{n-1}}{(1-R_o^{s+1})^{n+1}} \nonumber  \\  &=& r^2
(1+r)^{s+n-1} \frac{(1-(1+r)^s)^{n-1}}{(1-(1+r)^{s+1})^{n+1}}.
\end{eqnarray}
Again, the last result is valid for $n \geq 1$, while for $n=0$ one
obtains $P^{(s)}_0 = [1-R_o^s]/[1-R_o^{s+1}]$.

\section{Abundance of a clone as a function of its age}

Armed with the formula obtained in the last section, one can get
some basic intuition regarding the age-abundance relationships. The
\emph{average} abundance of a species $s$ generation after point
speciation (conditioned on non-extinction, i.e., the average number
of descendants of a single individual after $s$ generation, given
that this individual has at least one living descendant at this
time) is,
\begin{equation} \label{eq5z}
{\bar n(s)} = \frac{1}{1-P_0^{(s)}} \sum_{n=1}^\infty n P_n^{(s)} =
\frac{R_o^{s+1}-1}{R_o-1}.
\end{equation}

\begin{widetext}
\begin{figure}
\includegraphics[scale=0.4]{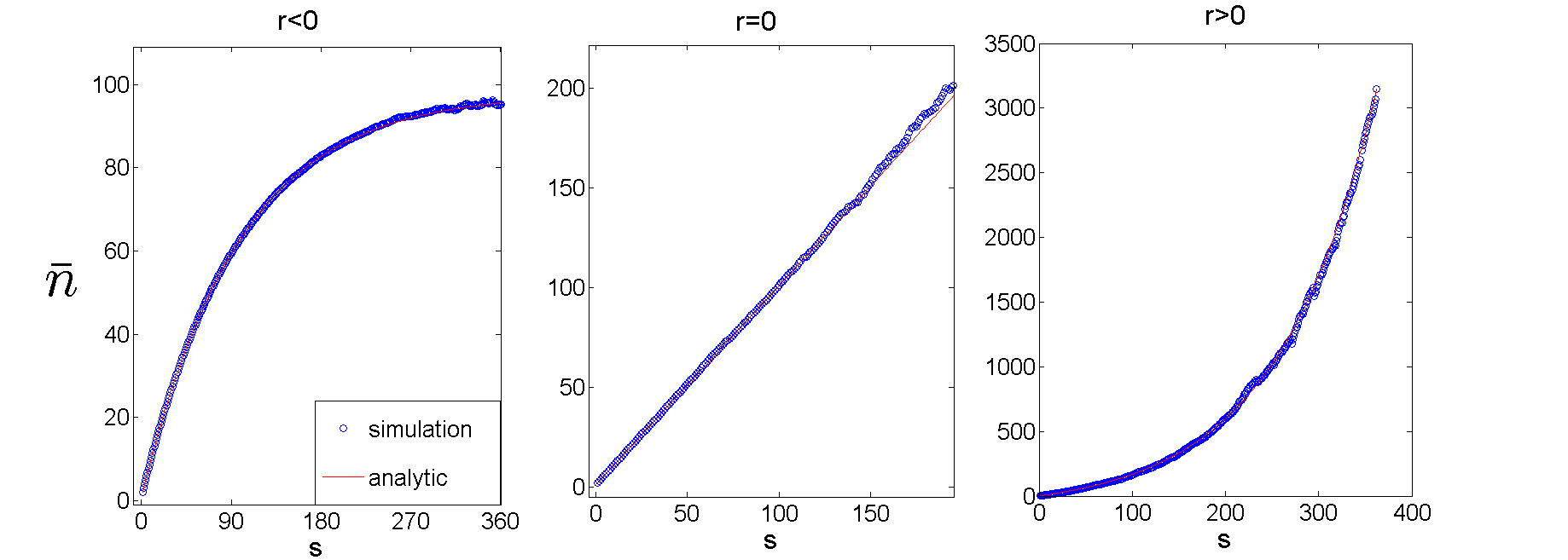}
\caption{The average size of a (non-extinct) species $s$ generation after its origination, ${\bar n(s)}$, as a function of $s$.
... present the results of a numerical simulation, while the full line is the analytic expression, Eq.  \ref{eq5z}. In the left panel,
$R_0 <1$ and $r<0$ meaning that, although the overall size of the community is growing $\gamma = {10}^{-2}$,
the abundance of every species is shrinking on
average as the demographic losses due to mutations $\nu = 2*{10}^{-2}$ are stronger. In that case, even species that survive
for a long time cannot cross (on average) the abundance $1/|r|$. In the middle panel the behavior of ${\bar n(s)}$
marginal case $r=0$ ($\gamma = {10}^{-2}, \ \nu = 0.99*{10}^{-2}$) is
graphed, where now the growth is strictly linear, while in the left panel ($\gamma = 2*{10}^{-2}, \ \nu = 0.0102$) the crossover to
exponential growth happens, again, at $1/r$. Overall, the growth is controlled by the "deterministic" parameter $r$
after $1/|r|$ generations, and before this stage the growth is linear, independent of $r$, and is related to the purely stochastic
dynamics at $r=0$. The numerical results presented here were obtained by simulation a Wright-Fisher  process with the appropriate values
of $\gamma$ and $\nu$, starting from a single individual and averaging over ${10}^8$ trials.         } \label{fig1}
\end{figure}
\end{widetext}

To understand the implications of (\ref{eq5z}), let us start with
the marginal case $R_o=1$ (or  $r=0$). In this case the overall
community is growing ($\gamma > 0$) but the average size of a single
clone is kept fixed on average, as the effect of growth  is balanced
by mutations. In this limit the average abundance of a surviving
clone is growing linearly with its age $s$, ${\bar n(s)}=s+1$. On
the other hand, the average abundance of all clones (including those
who went extinct and has abundance zero)  is time independent, since
the chance of surviving until $s$ scales like $1/s$. This feature is
demonstrated in the middle panel of Fig \ref{fig1}.

The marginal case $R_o=1$ is, of course, special and requires a
fine-tuning balance between mutations and growth rates. Still, in many
common situations the relevant values of both $\gamma$ and $\nu$ are
much smaller than unity (see, in example, the cases considered in
\cite{maruvka2011birth}), hence $r<<1$. In these cases, as long as
$|r|s<<1$ the surviving clones are still growing linearly in time as
in the marginal case. This linear pattern breaks down for $|r|s>1$:
if $r>0$ the clone switches to  exponential growth, ${\bar n}(s)
\sim \exp(rs)$ (Fig. \ref{fig1}, right), while for $r<0$,  ${\bar
n}(s)$ saturates to $1/r$  at large $s$, as demonstrated in the left
panel of Fig. \ref{fig1}. This latter case includes the scenario
considered in Hubbell-Kimura neutral theory, i.e., a fixed size
community, $\gamma = 0$, and finite chance of mutation, $\nu>0$.

\section{Age-abundance relationships: a maximum likelihood approach}

As seen in the last section, our solvable model allows us to
calculate the average abundance of a species given its age. However,
many practical problems have to do with the opposite question: given
the abundance of a species or a clone and its demographic features
represented by $r$, what is the best estimate one may suggest for
its age.

Perhaps the simplest approach towards this problem is to implement
Eq. (\ref{eq19}) and to calculate, given  $r$, the value of $s_{ML}$
that gives the maximum likelihood for $n$.
 It turns out that,
\begin{equation} \label{ml}
{\tilde s_{ML}} = \left\{   \begin{array}{cc}
                             n/2 & n|r| \ll 1 \\
                              \frac{ln(n|r|)}{|r|} & n|r| \gg 1
                           \end{array}  \right.
\end{equation}
Note that the expression for $s_{ML}$ is independent of the sign of
$r$, i.e., it takes the same value for growing and shrinking clones.
This feature is obvious in the linear regime $ n|r| \ll 1$, where
the system still behaves as if it is at criticality. Quite
surprisingly, however, even when $n|r| \gg 1$ $s_{ML}$ is still independent of the sign of $r$
despite the pronounced difference in the actual behavior of the two systems.
When $r>0$ the long term growth is
exponential, so the time to reach abundance $n$ is logarithmic in
$n$. When $r$ is negative, as demonstrated in Fig \ref{fig1}, clones
with abundance $n>1/|r|$ are exponentially rare and do not
contribute to ${\bar n}$. Still, the most likely course of events
that leads to this rare scenario is a rapid exponential growth, so
the age-abundance relationships are still logarithmic.

For the special case of a fixed size community
($\gamma = 0, \ \nu>0, \ r = -\nu <0$)  the species abundance
distribution (SAD) is known to be the Fisher log-series $exp(-\nu
n)/n$. Accordingly, species with $n>1/\nu$ are exponentially rare,
and typically the range of possible sizes of species is limited by
$1/|r|$. As followed from (\ref{ml}), \emph{if} such a rare event
occurs the age of the rare species scales with the logarithm of its
abundance (as opposed to the linear scaling for species in the bulk
of the SAD). The relatively young age of the exceptionally abundant
species cannot solve the species-age problem, since the chance to
observe species with $n>>1/\nu$ is tiny.

\section{The average age of species of a certain abundance}

In this section we would like to calculate the average age of a
species of abundance $n$ (as opposed to the maximum likelihood
method implemented in the last section). When the distribution
$P(s|n)$ is sharply peaked,  the mean and the maximum likelihood are
almost identical. On the other hand, when the distribution is wide
and skewed the average provides additional information and
facilitates the estimation based on finite number of data points.

To begin, we assume that the number of mutations/speciations per
generation is given by $\nu J(s)$, where $J$ is the overall size of
the community. When the size of the community is growing
exponentially one may assume further that $J(s) = exp(\gamma s)$.
Given that, the expression $P(n|s)$ (the chance that a species,
observed at a size $s$, was originated $n$ generations ago) may be
obtained from $P_n^{(s)}$ using elementary Bayesian inference
\cite{chisholm2014species},
\begin{equation} \label{eq13}
P(s|n) = \frac{P^{(s)}_n e^{-\gamma s}}{\sum_{s=1}^{T_{max}-1}
P^{(s)}_n e^{-\gamma s}}.
\end{equation}
The denominator in this expression is proportional to the  average
number of clones/species of size $n$ at the $s$ generation, a
quantity that was already calculated in the context of species
abundance distribution for growing and shrinking populations
\cite{manrubia2002boundary,maruvka2010universal,maruvka2011birth,maruvka2013model}.
Since $P^{(s)}_n$ is known, $P(s|n)$ may be written
explicitly, and we will provide the appropriate expression at the
end of this section. For the moment let us  calculate analytically
the average lifetime of a species given its abundance as suggested
in \cite{chisholm2014species},
\begin{equation}
T^*(n) \equiv \sum s P(s|n).
\end{equation}

Implementing a fictitious parameter $\tilde{\gamma}$, $T^*(n)$ may
be written in a simpler form (this step is analogous to the
derivation of the free energy from the partition function in
equilibrium statistical mechanics),
\begin{equation} \label{c1}
T^*(n) = \frac{\sum_{s=1}^{T_{max}-1} s P^{(s)}_n e^{-\gamma
s}}{\sum_{s=1}^{T_{max}-1} P^{(s)}_n e^{-\gamma s}} = \left(
-\frac{\partial}{\partial \Tilde{\gamma}} ln \left[
\sum_{s=1}^{T_{max}-1} P^{(s)}_n e^{-\Tilde{\gamma} s} \right]
\right)_{{\tilde \gamma}=\gamma}.
\end{equation}
Note that $P^{(s)}_n$ is also  $\gamma$ dependent, as shown in the last
section, via  $r = (1+ \gamma)(1-\nu)$ (or, if both rates are much
smaller than one, $r \approx \gamma - \nu$). To get the expression
in the r.h.s. it thus crucial to make a technical distinction
between the growth rate of the whole community ${\tilde \gamma}$ and
the growth rate of a single clone, $\gamma-\nu$ which is hidden in
$P^{(s)}_n$. As will be soon explained, this technical point facilitates the analytic calculation.

As mentioned, the argument of the log in the r.h.s. of (\ref{c1})
is, up to constants, $M_n$, the average number of species with
abundance $n$, which is simply the number of new species (mutants)
at a certain generation ($\nu$ times the number of birth which is
the size of the community at this generation) multiplied by the
chance that a mutant will generate a family of size $n$ when the
community is sampled,
\begin{equation}
M_n(t) = \sum_{s=1}^{t-1} \nu e^{\Tilde{\gamma} (t-s)} P^{(s)}_n.
\end{equation}
As shown in
\cite{manrubia2002boundary,maruvka2010universal,maruvka2011birth},
this quantity satisfies,
\begin{equation}
\frac{\partial M(n,t)}{\partial t} = \frac{\sigma^2}{2}
\frac{\partial^2}{\partial n^2} \left( nM(n,t) \right) +\left( \nu -
\gamma \right) \frac{\partial}{\partial n} \left( nM(n,t) \right).
\end{equation}

When the SAD reaches a stable distribution, the chance of a clone,
picked at random (independent of its abundance) from the list of
living clones, to have abundance $n$ is time-independent (this, of
course, cannot be true for the wild type as will be discussed soon).
Accordingly, for a growing community the number of $n$-clones,
$M_n$, must grow exponentially like $exp({\tilde \gamma} t)$. Hence,
\begin{equation}
\Tilde{\gamma}M(n) =  \frac{\sigma^2}{2} \frac{\partial^2}{\partial
n^2} \left( n M(n) \right) +\left( \nu - \gamma \right)
\frac{\partial}{\partial n} \left( n M(n) \right),
\end{equation}
yielding
\cite{manrubia2002boundary,maruvka2010universal,maruvka2011birth},
\begin{equation}
M(n) = \frac{A}{n} U \left( \frac{\Tilde{\gamma}}{| \gamma-\nu |}
,0,\frac{2| \gamma-\nu |}{\sigma^2}n \right)
\end{equation}
where $U$ is the Kummer function \cite{abramowitz1972handbook}.

Before plugging this expression into (\ref{c1}), we should consider
the normalization factor $A$. The normalization condition is,
\begin{equation} \label{nor1}
\int_0^\infty n M(n,s) dn = \int_0^\infty n M(n) e^{\tilde{\gamma}
s} dn = J_{\nu}(s),
\end{equation}
where $J_{\nu}(s)$ is the overall abundance of all the \emph{mutant}
species, i.e., the size of the community apart from the wild type,
\begin{equation}\label{nor2}
J_{\nu}(s) = \int_0^s \nu e^{\tilde{\gamma} s} e^{(\gamma-\nu)(t-s)}
dt = \frac{\nu e^{\tilde{\gamma} s}}{\tilde{\gamma}-\gamma+\nu}.
\end{equation}
Combining (\ref{nor1}) and (\ref{nor2}) one obtains:
\begin{equation}
\int_0^\infty n M(n) e^{\tilde{\gamma} s} dn = \frac{\nu
e^{\tilde{\gamma} s}}{\tilde{\gamma}-\gamma+\nu}.
\end{equation}
Plugging $M(n)$ into this expression one can find $A$, yielding,
\begin{equation}
M(n) = \frac{2}{\sigma^2} \frac{\nu}{n} \Gamma
\left(1+\frac{\Tilde{\gamma}}{| \gamma-\nu |} \right)  U \left(
\frac{\Tilde{\gamma}}{| \gamma-\nu |} ,0,\frac{2| \gamma-\nu
|}{\sigma^2}n \right).
\end{equation}
Finally, the average age of species with abundance $n$ is given by:
\begin{equation} \label{tstar}
T^*(n) = \left( -\frac{\partial}{\partial \Tilde{\gamma}} ln \left[
\Gamma \left(1+\frac{\Tilde{\gamma}}{| \gamma-\nu |} \right) U
\left( \frac{\Tilde{\gamma}}{| \gamma-\nu |} ,0,\frac{2| \gamma-\nu
|}{\sigma^2}n \right) \right] \right)_{{\tilde \gamma}=\gamma}.
\end{equation}
To compare with the maximum likelihood results presented above one
would like to translate $ \gamma-\nu $ to $r$. Note, however, that
the Fokker-Planck expression holds only for small values of $\gamma$
and $\nu$, i.e., when  $\gamma \nu$ is  negligibly small.

As above, we will consider two limits, $|r| n<<1$ and $|r| n>>1$.
Although the results are relevant for both positive and negative
$r$, our interest is mainly in the positive $r$ case so we will
explain the intuitive argument in the corresponding language. In the
first limit $r n<<1$ we deal with clones that are still in their
linear growth phase. Expanding  $T^*$ for small $rn$ one obtains:
\begin{equation}
T^*(n) \approx -n \left(\log (n r)+\frac{r}{\gamma }+\psi ^{(0)}\left(\frac{\gamma
}{r}\right)+\frac{\gamma  \psi ^{(1)}\left(\frac{r+\gamma
}{r}\right)}{r}+2 \gamma_E -1\right),
\end{equation}
so the leading term scales like $n \ln(n)$, as opposed to the linear
dependence obtained by maximum likelihood.

It is interesting to note
that, in the limit of fixed size community ($\gamma =0, \ r=\nu$),
\begin{equation}
T^*(n) \approx -n (\log (n \nu)+\gamma_E -1),
\end{equation}
and for large $n$ the identity $H_n \approx \gamma_E + ln(n)$ leads to
the result of \cite{chisholm2014species}:
\begin{equation}
T^*(n) = n \left[ 1- H_n - ln \left( \nu \right) \right].
\end{equation}

In the large $rn$ limit the species is in its exponential growth
phase so the age should be logarithmic in the abundance. Indeed, for
large $rn$ the Kummer function may be approximated by its power law
asymptote to yield,
\begin{equation}
\frac{\log (n r)}{r} + \frac{-H_{\frac{\gamma }{r}}+\gamma_E }{r},
\end{equation}
and the  leading term is identical with the maximum likelihood
estimation. Since our theory depends only on the absolute value of
$r$ one may take the limit $\gamma=0, \ |r|=\nu$ to obtain, again,
the result of \cite{chisholm2014species},
\begin{equation}
T^*(n) \approx \frac{1}{\nu} \left(\log (\nu n )+\gamma_E\right).
\end{equation}

Finally, since we have an expression for  $P_n^{(s)}$, the
the probability function  $P(s|n)$ may be derived from (\ref{eq13}).
For example, the geometric model yields,
\begin{equation}
P(s|n) = \frac{r^2(1+r)^{s+n-1}
\frac{(1-(1+r)^s)^{n-1}}{(1-(1+r)^{s+1})^{n+1}}}{\frac{2}{\sigma^2}
\frac{1}{n} \Gamma \left(1+\frac{\gamma}{| \gamma-\nu |} \right)  U
\left( \frac{\gamma}{| \gamma-\nu |} ,0,\frac{2| \gamma-\nu
|}{\sigma^2}n \right)}.
\end{equation}

\section{Discussion}

The main goal of this work was to extend the work of \cite{chisholm2014species} to communities with
finite growth (or decay) rates. To do that we used two approaches: the first was based on a solvable
model, the Wright-Fisher dynamics with geometric distribution of offspring per individual. This model allows for
an exact solution for the probability of a single individual to produce $n$ offspring after $s$ generations,
 $P_n^{(s)}$ (Eq. \ref{eq19}). The emerging picture is that the surviving species grow linearly in time as long as $s<1/|r|$;
 at longer times they either grow exponentially (if $r>0$) or saturate (if $r<0$). Eq. (\ref{eq19}) provides
  also a maximum likelihood
 estimation of the age of a species given its abundance. Quite surprisingly it turns out that the result
 does not depend on the sign of
 $r$. This apparently reflects some hidden symmetry  in this system.

 Our second approach was based on the Bayesian method suggested in \cite{chisholm2014species}, extending it to the case of growing
 and shrinking populations. In this case we have calculated $T^*(n)$, the average age of a species with abundance $n$. To do that we
 have implemented a trick, based on the distinction between the growth rate of a population $\gamma$ and the growth rate
 of the community ${\tilde \gamma}$. With this trick, the problem looks very much like the calculation of expectation values for
 a statistical physics system at equilibrium, and  the solution depends only on the species abundance distribution of the
 model. For growing or shrinking communities the SAD has already been calculated
  \cite{manrubia2002boundary,maruvka2010universal,maruvka2011birth}, and this led us to the expression (\ref{tstar}) and its
  low/high abundance asymptotics, including the results of \cite{chisholm2014species} as a special case.

  At this point, we would like to add two technical comments.

\begin{figure}
\includegraphics[width=9cm]{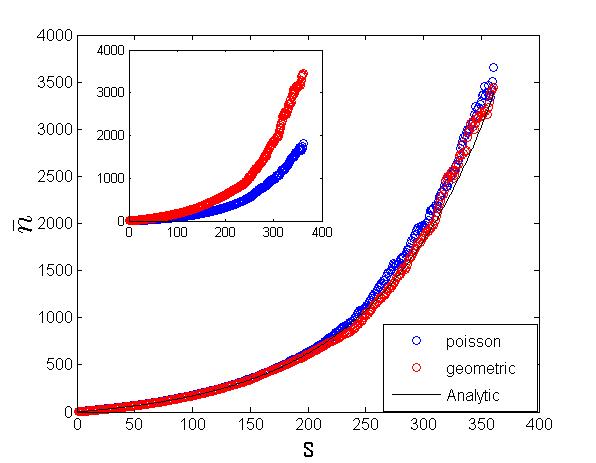}
\caption{${\bar n(s)}$ (conditioned on survival) for geometric (red) and Poisson (blue) distributions of the number of offspring per individual.
In the inset, the  number of offspring as a function of time, averaged over ${10}^8$ histories, is plotted for two communities. Both communities
have $\gamma = 2*{10}^{-2}$ and $\nu = {10}^{-2}$, and the only difference is the distribution function. Apparently, the geometric species grow much
faster; the reason is that the chance of extinction for a geometric species is higher, and the surviving species abundance should compensate it in order to preserve
the fixed mean. In the main panel both histories are plotted again, now the Poisson (red) points are divided by a factor of $\sigma^2/\sigma_0^2$,
where for Poisson $\sigma^2 = {(1-\nu)}^2e^\gamma + \nu R_0$ and ${\sigma_0}^2 = R_0(R_0+1)$  is the geometric distribution variance. Both datasets collapse, and fit the analytic expression \ref{eq5z}.    }
\label{fig2}
\end{figure}

\begin{enumerate}
  \item First, while the first approach was based on a specific model (geometric distribution
  for the number of offspring) for which
  the average number of same species offspring, $R_0 \equiv (1+\gamma)(1-\nu)$, determines the variance $R_0(1+R_0)$,
  the analysis of the second approach is based on the SAD, given in general terms without restriction
  to a certain relationships between the mean and the variance. The justification for that has to do with the
   \emph{universality} of the problem,
  as analyzed in detail in \cite{maruvka2010universal}.  Strictly speaking, the use of a
  Fokker-Planck equation is justified only in
  the limit of small $r$ and ${\cal O}(1)$ variance  $\sigma^2$,
  such that $P_n$ may be considered as a continuous function of $n$ and
  $s$. Therefore, given any distribution of offspring $P_n$ with mean $1+\gamma$ and variance $\sigma^2$,
  one can implement the Fokker-Planck equation to obtain $\tilde{P}^{(s)}_n = \alpha P^{(s)}({\alpha n})$
  where $P^{(s)}({n})$ fits our result (Eq. \ref{eq19}) and $$\alpha \equiv \frac{R_0(1+R_0)}{{(1-\nu)}^2 \sigma^2 + \nu R_0}.$$
  This feature is demonstrated in Fig. \ref{fig2} for geometric and Poisson offspring distributions.
  \item Another comment has to do with the applicability of the species abundance distribution for a growing community.
The SAD that we have analyzed is based on a \emph{steady state} solution of a Fokker-Planck equation. However, while
for a fixed size community, when the maximum abundance of a species scales with $1/\nu$, the equilibration time is also finite,
for a growing community  with $r>0$
there is no finite limit to the abundance and the equilibration time is always too small for the first few species (like
the wild type and the first mutants). Accordingly, the Kummer-like SAD analyzed here is inappropriate for the most abundant
species, like those that were considered by \cite{antal2011exact,kessler2013large}. Note, however, that the width of the distribution
for abundance values for these species is very wide, making a reliable inference of the age of the species from its size quite
difficult.
\end{enumerate}

Finally, we would like to discuss the applicability of a model with growth/decay of the population to the
species-age problem that arises in the context of Hubbell's neutral model. As we have already seen, the typical time
needed for a species to reach abundance $n$ is ${\cal O}(n)$  generations as long as $nr<1$, which is always the case when $\gamma=0$ and
the largest abundance in the system is  ${\cal O}(1/\nu)$, the regime considered in the neutral theory of biodiversity.
If, for example, one
considers a set of $10^9$ conspecific trees for a frequent species
in the Amazon basin (this is close to the contemporary figure, see
recent survey in \cite{ter2013hyperdominance}), with about a 50y
generation time, the neutral theory suggests $50 \cdot 10^9$
generations, more than the age of the universe.
 The tempo of the dynamics may be accelerated if one
assumes a very large value of $\sigma^2$ (as suggested, essentially,
in \cite{allen2007setting}, see \cite{chisholm2014species}. Note that the $n$ scaling is renormalized by $\sigma^2$ as we discussed
in the context of universality) or by
keeping the generation time as a free parameter (see, e.g.,
\cite{azaele2006dynamical}), but any of these approaches carries its
own difficulties when compared with realistic timescales and variances in empirical systems.

One of the original aims of this project was to check if one can solve this timescale problem by assuming that the trees in the
Amazon basin, say, are a \emph{growing} community. In many models evolution takes place as species increase their carrying capacity
(K-selection), so at least in principle it is possible that the population was much smaller in previous times. However, when we tried
to apply our analysis to the estimations suggested by \cite{ter2013hyperdominance} we failed to find an appropriate regime of parameters.
Specifically, a fit of Kummer function to the observed SAD result in very small values of $\gamma$, values that cannot
solve the age-abundance problem.

\bibliography{references_matan}

\end{document}